\documentclass[aps,prl,showpacs,superscriptaddress,twocolumn]{revtex4}

\usepackage{hyperref}
\usepackage{bbm}
\usepackage{graphicx}
\usepackage{amsmath}
\usepackage{amssymb}
\usepackage{mathscinet}
\usepackage{amsfonts,amsmath}
\usepackage{color}

\def\kb#1#2{|#1\rangle\!\langle #2 |}

\newcommand{\code}{\mathcal{C}}
\newcommand{\m}{\mathcal }

\usepackage[matrix,frame,arrow]{xypic}
\input{Qcircuit}

\begin{document}

\title{Using concatenated quantum codes for universal fault-tolerant quantum gates}

\author{Tomas Jochym-O'Connor}

\affiliation{Institute for Quantum Computing, University of Waterloo, Waterloo, Ontario, N2L 3G1, Canada}
\affiliation{Department of Physics \& Astronomy, University of Waterloo, Waterloo, Ontario, N2L 3G1, Canada}

\author{Raymond Laflamme}

\affiliation{Institute for Quantum Computing, University of Waterloo, Waterloo, Ontario, N2L 3G1, Canada}
\affiliation{Department of Physics \& Astronomy, University of Waterloo, Waterloo, Ontario, N2L 3G1, Canada}
\affiliation{Perimeter Institute for Theoretical Physics, 31 Caroline Street North, Waterloo, Ontario, N2L 2Y5, Canada}

\date{\today}

\begin{abstract}
We propose a method for universal fault-tolerant quantum computation using concatenated quantum error correcting codes. Namely, other than computational basis state preparation as required by the DiVincenzo criteria~\cite{DiVincenzo00}, our scheme requires no special ancillary state preparation to achieve universality, as opposed to schemes such as magic state distillation. The concatenation scheme exploits the transversal properties of two different codes, combining them to provide a means to protect against low-weight arbitrary errors. We give the required properties of the error correcting codes to ensure universal fault-tolerance and discuss a particular example using the 7-qubit Steane and 15-qubit Reed-Muller codes. We believe that optimizing the codes used in such a scheme could provide a useful alternative to state distillation schemes that exhibit high overhead costs.
\end{abstract}


\maketitle

\section{Introduction}
The ability to physically manipulate quantum mechanical systems promises to provide a means towards powerful quantum computing and simulation~\cite{Shor97, Feynman82, Lloyd96}. Understanding and controlling sources of noise during the manipulation of quantum systems is fundamental towards the development of scalable devices that could achieve such computing promises. The theory of quantum error correction has been developed to address the latter, protecting quantum systems through the use of additional ancillary systems. Quantum error correction has had a rapid progression to address multiple types of errors and situations, and provides a building block to large scale quantum devices using fault-tolerant quantum computation. 

The goal of fault-tolerant quantum computation is to control quantum errors in a coherent way such that they do not propagate throughout the different quantum systems that are being coupled for the use of quantum computation. Any two-qubit coupling gates can propagate errors and are typically avoided as multiple errors may lead to logical faults after the application of quantum error correction. However, in order for such schemes to provide universal quantum computation, additional resources are required, typically through the preparation of special quantum states~\cite{Shor96, AB97, Knill05, BK05}. Addressing quantum noise in this manner allows for the establishment of noise thresholds, levels of noise for which scalable quantum computation is achievable without exponential overhead in resources~\cite{Shor96, Preskill98, KLZ98, AB97, Knill05}. In certain cases, rigorous numerical values of the threshold have been established by calculating the exact propagation of errors given a fixed error model and method of encoding for logical computation~\cite{AGP06, AGP08, PR12}.

One of the most widely used methods for fault-tolerant quantum computation is magic state distillation~\cite{BK05}, which promotes transversal Clifford gate operations to universal quantum computation through gate teleportation. While providing a means to increase the fault-tolerance threshold, the overhead in the preparation scheme for magic state distillation remains one of the large bottlenecks for scalable quantum computing, estimated to account for up to 90\% of the overall number of qubits in certain architectures~\cite{FMMC12}. As such, much effort is being invested into understanding and reducing the overhead associated with such schemes~\cite{MEK12, Jones12, BH12, JYHL13}. While such research has paved the way for the reduction of the overall cost of fault-tolerant quantum computation, this work will take a different approach, by using concatenated quantum error correcting codes to provide universal fault-tolerance, rather than state distillation. The scheme we propose uses two different quantum error correcting codes in concatenation. We argue that by sacrificing the full distance of the concatenated quantum error correcting code, we can exploit the transversal properties of both quantum codes to produce a set of operations that, while not globally transversal, provide a means for universal fault-tolerant quantum gates. In this work we shall focus on protecing against arbitrary single-qubit errors, however, we provide a brief description of how such a scheme could be generalized to correct against $t$~errors. Recently, Paetznick and Reichardt have proposed a similarly motivated work on universal quantum fault-tolerance without the preparation of special ancillary states, yet our work differs in the method of execution of the fault-tolerant logic~\cite{PR13}. Additionally, there has been research that has focused on obtaining a set of fault-tolerant operations to transfer between different quantum error correcting codes, yet such schemes do not yield a set of universal operations~\cite{SEDH08, HFWH13, Criger13}.

The outline of the work is as follows: we shall begin by introducing the necessary framework for universal gate sets and transversal logic. We follow by describing the requirements for the concatenated scheme. We then describe the method for implementing universal fault-tolerant logic and quantum error correction and conclude with a specific example and discussion.

\section{Preliminaries}
In order to fully exploit the power of quantum computers, simulating any completely positive mapping of quantum states is essential. By Stinespring's dilation theorem, any such mapping can be simulated by unitary transformations~\cite{Stinespring55}. As such, having the ability to approximate arbitrary unitary transformations is vital to implementing quantum algorithms and simulations.

Let $\m{G}$ be a finite set of unitary operators on a Hilbert space~$\m{H}$. We say that the set~$\m{G}$ is a universal gate set on~$\m{H}$ if any unitary transformation~$U$ in~$\m{H}$ can be approximated efficiently using gates from the gate set~$\m{G}$, that is given a target fidelity~$\epsilon$ the unitary~$U$ can be approximated using~$O(\log^c{1/\epsilon})$ gates from the gate set~$\m{G}$~\cite{Kitaev97, DN06}. Given a Hilbert space~$\m{H}$ composed of multiple qubits, any universal set of quantum gates on each of the individual qubits along with entangling gates coupling the qubits form a universal gate set for the full Hilbert space~$\m{H}$~\cite{BBC+95}. The universal gate set that we shall focus on in this work's example will be the set~$\m{G} = \{ H, T, CNOT \}$, where $H$ is the Hadamard gate, $T$ is the $\pi/8$-gate ($T = e^{-i\pi/8} \kb{0}{0} + e^{i \pi/8} \kb{1}{1}$ ), and $CNOT$ is the two-qubit controlled-not gate~\cite{BMPRV99}. 

Researchers focus on universal gate sets since developing techniques to deal with errors associated with a finite set of gates is a much more tractable task than correcting for faults for arbitrary unitary gates. As such, quantum error correcting codes are constructed to best protect against errors in the implementation of logical gates for a chosen universal gate set. We shall denote the weight of an error as the number of locations where a given error acts non-trivially (not the identity). One of the simplest methods to construct fault-tolerant schemes is by applying gates transversally. A logical gate~$g$ on a quantum error correcting code~$\m{C}$ is called $t$-transversal if $g$ interacts with at most $t$~locations of the underlying qubits composing the code~$\m{C}$. Unfortunately, it has been shown that no quantum error correcting code contains a universal set of transversal gates~\cite{EK09, ZCC11}. This motivates the search for different fault-tolerant methods to implement universal quantum logic.

\section{Concatenated QEC}
We propose using concatenated quantum error correction to achieve a set of universal fault-tolerant quantum gates. The idea behind such a scheme is that we can use the properties of one code to protect against the disadvantages of the other, and vice versa.  

The general concatenated error correcting scheme is as follows: the qubits that we desire to protect errors against are encoded into a quantum error correcting code~$\code_1$. In this work, we shall require the code distance of~$\code_1$ to be at least 3, so that it can correct arbitrary single-qubit errors. The qubits that make up the code~$\code_1$ are subsequently encoded into a second code~$\code_2$, which again will be required to have distance of at least 3. As we are focusing on codes that correct for an arbitrary single-qubit error, we shall refer to a transversal gate for a given code as a gate which is 1-transversal and any gate not having this form as non-transversal.


\vspace{-0.5cm}
\begin{figure}[h]
\centering
\[
\Qcircuit @C=1.3em @R=1.3em {
&& {/} \qw & \qw_{\m{C}_2} & \qw & \qw & \gate{g_2} & \qw \\
&& {/} \qw & \qw_{\m{C}_2} & \gate{g_2} & \qw & \qw & \qw \\
\m{C}_1 && \vdots & & & & & \\
&& {/} \qw & \qw_{\m{C}_2} & \qw & \qw & \gate{g_2} & \qw \\
&& {/} \qw & \qw_{\m{C}_2}  & \gate{g_2} \qwx[-3] & \qw & \qw & \dstick{g_1}  \qw 
 \gategroup{1}{2}{5}{2}{1.0em}{\{}  \gategroup{1}{5}{5}{7}{1.4em}{--}
}
\]
\caption{General construction of a logical gate for a concatenated error correction scheme. The qubit of information is encoded in a quantum error correcting code~$\code_1$, whose qubits are in turn encoded into a code~$\code_2$. As such, the logical gate~$g_1$ (given by the boxed region) on the encoded space~$\code_1$ will be composed of multiple logical gates $g_2$ on the $\code_2$~codeblocks.}
\label{fig:logicalT}
\end{figure}
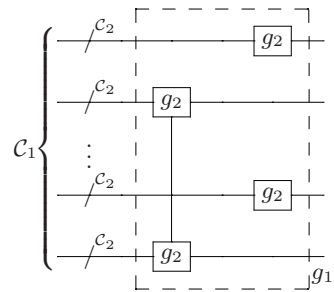

The important properties for the quantum error correcting codes~$\code_1$ and~$\code_2$ for the implementation of universal fault-tolerant quantum logic are as follows:
\emph{1. }For any logical gate that is non-transversal in~$\code_1$, there must exist an application of this logical gate using gates that are transversal in~$\code_2$, \emph{2. }The recovery operations (syndrome measurement and error correction operations) on $\code_1$ and $\code_2$ must be globally transversal (in the full concatenated code space).


\section{Implementing universal fault-tolerant logic}
\subsection{Logical Gates}
We first explain the requirements of the set of gates to implement a universal set of fault-tolerant operations. Since there exists no quantum error correcting code that exhibits a full set of transversal quantum gates~\cite{EK09, ZCC11}, there will always be at least one gate in a given universal gate set that will couple qubits that make up the error correcting code, leading to the possibility of bad error propagation. Consider the first level of encoding~$\code_1$, the non-transversal gate can lead to a propagation of errors, however, if we are now encoding each of the qubits making up the code~$\code_1$ into a further error correcting code, the propagation from a single to multiple physical faults will not necessarily lead to a propagation of logical faults if the errors are sufficiently sparse.

Specifically, the first requirement of the concatenated quantum error correction scheme stipulates that every non-transversal gate in the code~$\code_1$ can be implemented using transversal gates in the code~$\code_2$. The non-transversality of a given gate will cause the propagation of a single physical fault between different logical qubits in~$\code_1$. The implementation of the non-transversal~$\code_1$ gates will govern the propagation of the physical errors between the qubits. Therefore, we require the gates that make up the logical gate on~$\code_1$, themselves logical gates for the code~$\code_2$, to be transversal in the encoded space~$\code_2$. By imposing such a restriction, a single error occurring in the non-transversal gate application in~$\code_1$ will propagate to at most a single physical error in each of the logical qubits forming~$\code_1$, which themselves are encoded blocks of~$\code_2$. This is precisely the property one is after in a fault-tolerant quantum computation, that a single physical error will propagate to at most a single physical error on encoded codeblocks, allowing for the correction of such errors.

Given a choice of codes~$\code_1$ and $\code_2$, not all gates of the universal gate set will be transversal in~$\code_2$. By the properties outlined above, any logical gate in~$\code_1$ that uses gates from~$\code_2$ that are not transversal in its construction, must be transversal in~$\code_1$. In performing such a gate, a single fault on a particular $\code_2$~codeblock could propagate to multiple errors within this codeblock and could lead to a logical $\code_2$~fault in the codeblock where the error occurred. However, a single logical fault on one of the $\code_2$~codeblocks will not yield a global logical fault on~$\code_1$, as such a code can correct for arbitrary logical faults on one of its encoding logical qubits. 


\subsection{Quantum Error Correction}
How is error correction then applied? We shall describe the error correction properties that are required after the application of two types of logical quantum gates, those that are non-transversal in~$\code_1$ yet use an application of transversal~$\code_2$ gates, and the application of logical gates that are transversal in~$\code_1$, whose individual block gates are non-transversal in~$\code_2$. In the case of the former, the important property of the error correction is that it does not couple qubits within the codeblocks of~$\code_2$, as the application of the gate could propagate a single fault into multiple single faults on each of the~$\code_2$ codeblocks. If the error correction procedure propagates errors within the $\code_2$~codeblocks, then single errors on each codeblock will propagate to multiple errors on each codeblock, thus possibly leading to logical errors on multiple codeblocks, therefore causing a global logical fault. As such, it is very important that the error syndrome measurement and correction be performed transversally on each of the $\code_2$~codeblocks. Error correction at the $\code_1$ level is not necessary after the application of this type of logical gate, as the errors propagate within the codeblocks and the scheme is constructed in a way that all such errors on the codeblocks are recoverable as long as only a single fault occurs in the application of the gate.

The error correction procedure after the implementation of the transversal gate in~$\code_1$ (using non-transversal $\code_2$ gates) will require an additional level of error correction. As in the application of the non-transversal $\code_1$~gates, error correction on each of the $\code_2$ codeblocks is first applied. As the application of the logical gate on~$\code_1$ uses non-transversal $\code_2$~gate applications, a single (correctable) error on a particular $\code_2$~codeblock can propagate to a non-correctable set of errors on that given codeblock. As such, performing the $\m{C}_2$~error correction on that codeblock will introduce a logical error (if the error were to occur during the $\code_2$ error correction process itself then this error will be weight one). 
However, as mentioned above, if only a single logical $\m{C}_2$~error has occurred, the logical fault introduced by the error correction will be correctable using error correction procedure on~$\code_1$. However, it is important that the error correction procedure on~$\mathcal{C}_1$, which is a logical error correction procedure as it acts on logically encoded states in~$\mathcal{C}_2$, is itself globally transversal. As such, errors that could occur during error correction will not propagate to multiple physical errors that could be detrimental upon the application of further logical computation.


\section{Example: A [[105,1,3]] Quantum Error Correcting Code}

A simple example of the scheme outlined in this work involves two of the most well studied quantum error correcting codes, $\code_1$ will be the 7-qubit Steane code~\cite{Steane96b} and $\code_2$ the 15-qubit Reed-Muller code~\cite{KLZ96}. The 15-qubit Reed-Muller code has the following set of transversal gates: $\{T, CNOT \}$, where each logical gate is achieved by applying the gate itself to each of the qubits (or $T^{\dagger}$ in the case of the logical $T$~gate). The missing gate from the universal gate set is the Hadamard gate. The 7-qubit Steane code (corresponding to~$\code_1$) has the following set of transversal gates: $\{S, H, CNOT\}$, where $S = \kb{0}{0} +i \kb{1}{1}$.  Each logical gate is achieved by applying an individual gate to each of the qubits, or pair of qubits (in the case of applying logical $S$, one applies $S^{\dagger}$ to each of the qubits). As such, $\code_1$ can implement gates from the Clifford group transversally, yet are missing the $T$~gate from the universal gate set that can be implemented transversally. 

The concatenated code is 7 blocks of 15 qubits, totalling 105 qubits, encoding 1 qubit of information. As both quantum codes share the property that all Pauli gates, the $S$ phase gate, and the $CNOT$ gate can be implemented logically by applying the gate to each qubit, or pair of qubits, then the globally logical version of these gates for the 105 qubit code are also achieved by applying the corresponding gate to each qubit, or pair of qubits, of the full 105 qubit code. Additionally, all syndrome measurements (which will correspond to the measurement of Pauli observables) will be transversal within the code, as well as the Pauli corrections.

\vspace{-0.5cm}
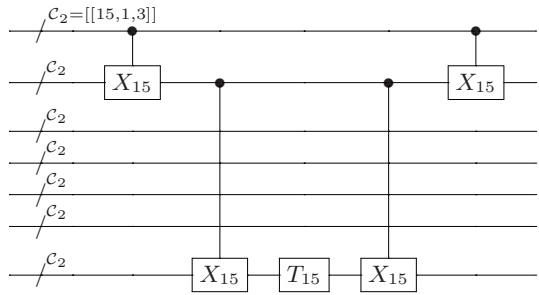
\begin{figure}[h]
\centering
\[
\Qcircuit @C=1.3em @R=1.3em {
& {/} \qw & \qw & \ctrl{1}_{\m{C}_2 = [[15, 1, 3]]} & \qw & \qw & \qw & \ctrl{1} & \qw \\
& {/} \qw & \qw_{\m{C}_2} & \gate{X_{15}} & \ctrl{5} & \qw & \ctrl{5} & \gate{X_{15}} & \qw \\
& {/} \qw & \qw_{\m{C}_2} & \qw & \qw & \qw & \qw & \qw & \qw \\
& {/} \qw & \qw_{\m{C}_2} & \qw & \qw & \qw & \qw & \qw & \qw \\
& {/} \qw & \qw_{\m{C}_2} & \qw & \qw & \qw & \qw & \qw & \qw \\
& {/} \qw & \qw_{\m{C}_2} & \qw & \qw & \qw & \qw & \qw & \qw \\
& {/} \qw & \qw_{\m{C}_2}  & \qw & \gate{X_{15}} & \gate{T_{15}} & \gate{X_{15}} & \qw & \qw
}
\]
\caption{Logical $T$ gate for the Steane code~$\code_1$, composed of logical $CNOT_{15}$ and $T_{15}$ gates on the $\code_2$ codeblocks. These gates are transversal in $\code_2$, and therefore only propagate errors to different codeblocks, without propagating within a given $\code_2$ codeblock.}
\label{fig:logicalT}
\end{figure}

The logical $T$ gate is achieved by combining logical gates on the different $\code_2$~codeblocks, which as shown in Figure~\ref{fig:logicalT} is not transversal in the $\code_1$~code, yet uses gates that are all transversal within the 15 qubit $\code_2$~codeblocks. As explained in the previous section, a single error in the implementation of the logical gate can propagate to multiple errors (a maximum of 3 for this particular gate application) yet will be distributed such that there are only single errors on each~$\code_2$ codeblock. The error correction procedure measures the syndromes on each of the $\code_2$~codeblocks individually, which corresponds to measuring the 14 syndromes corresponding to the 15-qubit code. The Pauli error correction operations are then applied to correct for the errors that occurred during the application of the logical $T$ gate. As such, the concatenated code can correct for an arbitrary weight-1 error that occurs during the implementation of the logical~$T$ gate. 

\begin{figure}[h]
\centering
\[
\Qcircuit @C=1.3em @R=1.3em {
& {/} \qw & \qw_{\m{C}_2} & \gate{H_{15}} & \qw \\
& {/} \qw & \qw_{\m{C}_2} & \gate{H_{15}} & \qw \\
& {/} \qw & \qw_{\m{C}_2} & \gate{H_{15}}  & \qw \\
& {/} \qw & \qw_{\m{C}_2} & \gate{H_{15}}  & \qw \\
& {/} \qw & \qw_{\m{C}_2} & \gate{H_{15}}  & \qw \\
& {/} \qw & \qw_{\m{C}_2} & \gate{H_{15}}  & \qw \\
& {/} \qw & \qw_{\m{C}_2}  & \gate{H_{15}}  & \qw
}
\]
\caption{Logical $H$ gate for the Steane code~$\code_1$, implemented using logical $H_{15}$ gates on each of the $\code_2$ codeblocks. Note that the individual $H_{15}$ are non-transversal on each codeblock.}
\label{fig:logicalH}
\end{figure}
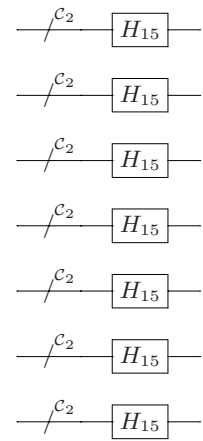

In order to implement the logical~$H$, one applies the logical~$H_{15}$ on each of the $\code_2$ codeblocks, as such it is transversal in the encoded states that form the code~$\code_1$, yet each individual $H_{15}$ is not transversal in its implementation on the $\code_2$ codeblocks. A single error that occurs in one of the individual applications of the $H_{15}$ gates could propagate to multiple errors within the codeblock, leading to possible logical errors. However, if only one such error occurs the full quantum code will still be protected. After the action of the gate, error correction is applied to each of the $\code_2$~codeblocks, possibly resulting in correction causing a $\code_2$ logical faults. However, if only one such logical $\code_2$~fault occurs, subsequent global error correction at the logical $\code_1$~level will detect such an error. The $\code_1$ error correction involves measuring the 6~stabilizers of the 7-qubit Steane code, where each stabilizer is now a logical stabilizers composed of $X_{15}$ or $Z_{15}$ operators. However, as such operators are transversal for the 15-qubit code, they can be measured in a transversal way. The maximal weight of the stabilizers measured for~$\code_1$ code is 32, since each of the logical $X_{15}$ gates involve 8~$X$~gates on the $\code_2$ codeblock, and the weight of the 7-qubit $X$ stabilizers is 4. Error correction for the $\code_1$~level will then be completed by performing logical Pauli error correcting operations on affected $\code_2$~codeblocks. The measurement of the $\code_1$ stabilizers will be the most expensive error correction step due to the high weight of the stabilizers. 

As described, the concatenated code can correct for any weight one error. However, it is worth noting that if one used a straight concatenation of the two codes to protect against quantum noise, the concatenated code will be a [[105,1,9]]~quantum error correcting code, that is, it would protect against 4 arbitrary errors. In this fault-tolerant scheme, we are sacrificing the larger distance of a straight concatenation scheme in order to protect against arbitrary single qubit errors when performing logical gates. For example, when implementing the logical~$T$~gate, since different codeblocks are coupled by $\code_2$~transversal gates between the codeblocks, two errors on a given codeblock could propagate to two errors on multiple codeblocks. Such propagation is bad, as this could cause logical faults on multiple blocks, an irrecoverable error. Similarly, in the implementation of the logical $H$~gate, single errors on different codeblocks, totalling at least 2 independent errors, could propagate through the non-transversal $H_{15}$~gate that is applied to each codeblock to result in multiple logical codeblock errors, resulting in an overall logical fault after error correction. Therefore, while the algebraic distance of the code is 9, due to the coupling of the different qubits when implementing the logical gates, the code only strictly protects against single physical faults, acting as a logical distance 3 code.

\section{Generalizing to QECC correcting against $t$~errors}
The scheme described in this work could be adapted to account for quantum error correcting codes~$\m{C}_1$ and~$\m{C}_2$ that correct against arbitrary weight~$t$ errors. The key properties of universal gate sets developed for such a concatenation scheme would be modified such that given a gate which is not $t$-transversal in $\m{C}_1$, the logical gates in $\m{C}_2$ which form such a gate must be $t$-transversal in $\m{C}_2$ when applied in composition. Additionally, similar requirements for the quantum error correction operations would be necessary. The error correction operations should be~1-transversal as to not possibly propagate errors that occur during the error correction process to multiple errors that could be detrimental at the next stage of computation.


\section{Conclusion}
In this work we have proposed a method for universal quantum fault-tolerance using concatenated error correcting codes. The full distance of the concatenated scheme is sacrificed in order to establish a set of universal quantum gates that are robust to a smaller set of errors. The transversal properties of the two different error correction schemes are exploited to limit the propagation of errors to either be sufficiently sparse, only a small number of errors per encoded codeblock, or limiting all errors to be contained within a single codeblock. We provide an example of such a scheme using the well-studied 7-qubit Steane and 15-qubit Reed-Muller codes. Providing rigorous thresholds of such a scheme, using methods such as those presented in Ref.~\cite{AGP06}, remains an interesting problem, as well as establishing the performance of such a scheme using other quantum error correcting codes.

\section{Acknowledgements}
The authors would like to thank Ben~Criger, Daniel~Gottesman, and Adam~Paetznick for insightful discussions. T.~J.~is supported by the Ontario Ministry of Training, Colleges and Universities and the Fonds de recherche du Qu\'ebec -- Nature et technologies. This work is supported by Industry Canada, CIFAR, and NSERC.

\bibliographystyle{apsrev4-1}
\bibliography{bibtex_jochym}

\end{document}